\documentclass[aps,prd,preprint,nofootinbib,11pt]{revtex4}
\usepackage{amsfonts}
\usepackage{mathrsfs}
\usepackage{graphicx}
\usepackage{amsmath}
\usepackage{amssymb}
\usepackage{subfigure}
\usepackage{epsfig}
\usepackage{graphicx}
\usepackage{ulem}
\usepackage{color}
\usepackage{slashed}
\usepackage{hyperref}
\newcommand{\bqa}{\begin{eqnarray}}
\newcommand{\eqa}{\end{eqnarray}}
\newcommand{\beq}{\begin{equation}}
\newcommand{\eeq}{\end{equation}}
\allowdisplaybreaks[1]
\graphicspath{{fig/}{dia/}} \DeclareGraphicsExtensions{.eps}

\begin{document}
\title{\Large Nodal mechanism for the suppressed $D\bar D$ decay of $\psi(4040)$ in the Bethe--Salpeter framework\\[7mm]}

\author{Bing-Dong Wan$^{1,2}$\footnote{wanbd@lnnu.edu.cn} and Sheng-Qi Zhang$^3$\footnote{shqzhang@pku.edu.cn}}

\affiliation{$^1$School of Physics and Electronic Technology, Liaoning Normal University, Dalian 116029, China\\
$^2$ Center for Theoretical and Experimental High Energy Physics, Liaoning Normal University, Dalian 116029, China\\
$^3$ Center for High Energy Physics, Peking University, Beijing 100871, China
}
\author{~\\~\\}
\begin{abstract}
The strong decay $\psi(4040)\to D\bar D$ is anomalously suppressed despite ample phase space, whereas the $D\bar D^*$ and $D_s\bar D_s$ channels remain sizable. In this work, we study this suppression and the associated open-charm hierarchy in the framework of the instantaneous Bethe--Salpeter equation combined with the relativistic $^3P_0$ model, with the pair-creation strength fixed independently from $\psi(3770)\to D\bar D$. Within this framework, we show that the suppressed $D\bar D$ mode can be understood as a consequence of node-induced cancellations in the relativistic decay amplitude. The $D\bar D$ amplitude is strongly reduced because the corresponding overlap integral receives comparable positive and negative contributions from different momentum regions, whereas the $D\bar D^*$ and $D_s\bar D_s$ channels do not undergo the same strong cancellation. This interpretation is further supported by the pronounced sensitivity of the $D\bar D$ width to the initial mass, the charged-neutral $D$-meson mass splitting, and the dip structure in the mass dependence of the partial width. Our results provide a dynamical explanation of the suppressed $D\bar D$ mode and the core open-charm hierarchy of $\psi(4040)$ within a conventional $3\,{}^3S_1$ charmonium picture, while the precise value of the near-vanishing $D\bar D$ width remains model dependent.
\end{abstract}
\maketitle

\section{Introduction}

The strong decays of excited charmonium states provide important information on their internal structure and decay dynamics. Among them, the vector state $\psi(4040)$ exhibits a particularly intriguing feature: the $D\bar D$ channel is strongly suppressed despite ample phase space, while the $D\bar D^*$ channel dominates and $D_s\bar D_s$ remains sizable~\cite{PDG}. This hierarchy deviates markedly from naive phase-space expectations and has long been regarded as a nontrivial feature of excited-charmonium decays. In the same energy region, the nearby state $\psi(3770)$ was first observed in early $e^+e^-$ experiments around $3.77$ GeV~\cite{Rapidis:1977cv,Scharre1978} and has since been studied extensively through its open-charm decays~\cite{Ablikim2004,Ablikim2006,Besson2006}.

If $\psi(4040)$ is interpreted as a conventional $3\,{}^3S_1$ charmonium state, one would not expect such a strong suppression of the $D\bar D$ mode from kinematics alone. Various theoretical approaches, including quark-pair-creation models, potential models, and coupled-channel analyses, have been employed to study these decays~\cite{Micu,LeYaouanc,GodfreyIsgur,EichtenQuigg,EichtenLaneQuigg2004,Eichten:2005ga,LeYaouancBook}. Existing studies further show that a $3S$ assignment can reproduce the qualitative hierarchy, and that nodal effects play an important role in this context~\cite{Barnes2005,Eichten2019,Ferretti2018}. In most of those discussions, however, nodal effects mainly serve as a qualitative explanation for why the $D\bar D$ width may be small. What remains less explicit is the amplitude-level origin of the cancellation: which quantity changes sign, how the cancellation is distributed among Salpeter components, and why the same initial-state node suppresses $D\bar D$ but not $D\bar D^*$ or $D_s\bar D_s$.

For radially excited states, the decay amplitudes are expected to be sensitive to the nodal structure of the wave function. In a relativistic Bethe--Salpeter description, however, the node is distributed among several Salpeter components rather than represented by a single radial zero. The relevant question is therefore not whether one component vanishes, but whether the full channel-dependent overlap integral develops a strong cancellation. A quantitative answer requires a framework in which the relativistic bound-state wave function and the decay amplitude are treated consistently.

In this work, we study the suppressed $D\bar D$ mode of $\psi(4040)$ and the associated open-charm hierarchy within the instantaneous Bethe--Salpeter (BS) equation combined with the relativistic $^3P_0$ model~\cite{BetheSalpeter,Salpeter,Chang2005,Wang2006,Wang2007,Wang2013,Ding2021}. The pair-creation strength is fixed independently from the measured decay $\psi(3770)\to D\bar D$~\cite{PDG}, and is then applied to $\psi(4040)$ without further adjustment. Our aim is not merely to reproduce the hierarchy, but to identify its dynamical origin in the relativistic amplitude itself. More specifically, we show that the relevant nodal object is the complete channel-dependent overlap integrand, not any individual radial component, and that within the BS framework the suppressed $D\bar D$ mode can be traced to sign-changing cancellations in that integrand, whereas the same cancellation does not occur for $D\bar D^*$ or $D_s\bar D_s$.

This interpretation is further constrained by two correlated signatures. First, the $D\bar D$ width exhibits a pronounced dip as the initial mass is varied near the physical region. Second, the small charged-neutral $D$-meson mass splitting is strongly amplified only because the $D\bar D$ amplitude is already near zero. Taken together, these features support an amplitude-level and internally consistent explanation of the suppressed $D\bar D$ mode and the core open-charm hierarchy of $\psi(4040)$ within a conventional quarkonium picture.

This paper is organized as follows. In Sec.~II, we introduce the instantaneous Bethe--Salpeter formalism and the relativistic $^3P_0$ model used in the calculation. In Sec.~III, we present the input parameters, the determination of the decay strength from $\psi(3770)\to D\bar D$, and the resulting wave functions. In Sec.~IV, we give the numerical results for the open-charm decay widths of $\psi(4040)$ and demonstrate the nodal mechanism underlying the suppressed $D\bar D$ channel. Section~V contains the discussion and conclusions.

\section{Formalism}

\subsection{Instantaneous Bethe--Salpeter equation}

In this work, the open-charm strong decays of the vector charmonium state $\psi(4040)$ are studied in the framework of the instantaneous Bethe--Salpeter (BS) equation combined with the relativistic $^3P_0$ model. The BS equation provides the relativistic wave functions of the initial charmonium state and the final heavy-light mesons, while the OZI-allowed decay amplitudes are evaluated through the creation of a light quark-antiquark pair from the vacuum. Since the formalism is standard, we summarize only the ingredients needed for the later discussion of the channel-dependent overlap integrals.

The BS wave function of a quark-antiquark bound state with total momentum $P$ is defined in the standard way~\cite{BetheSalpeter,Salpeter,Chang2005}. In momentum space, the quark momenta are parameterized as
\begin{equation}
p_1=\alpha_1 P+q,\qquad p_2=\alpha_2 P-q,
\qquad
\alpha_1=\frac{m_1}{m_1+m_2},\qquad \alpha_2=\frac{m_2}{m_1+m_2}.
\end{equation}
The BS equation then reads
\begin{equation}
S_1^{-1}(p_1)\chi_P(q)S_2^{-1}(-p_2)
=i\int\frac{d^4k}{(2\pi)^4}V(P;q,k)\chi_P(k),
\end{equation}
where $V(P;q,k)$ is the interaction kernel.

Following the standard instantaneous approximation, the interaction kernel is assumed to depend only on the components transverse to the total momentum,
\begin{equation}
V(P;q,k)\Big|_{\mathbf{P}=0}\simeq V(q_\perp,k_\perp),
\end{equation}
where
\begin{equation}
q^\mu=q^\mu_\parallel+q^\mu_\perp,\qquad
q^\mu_\parallel=\frac{P\cdot q}{M^2}P^\mu,\qquad
q^\mu_\perp=q^\mu-\frac{P\cdot q}{M^2}P^\mu,
\end{equation}
and $M=\sqrt{P^2}$ is the meson mass. The corresponding three-dimensional Salpeter wave function is defined as
\begin{equation}
\varphi_P(q_\perp)=i\int\frac{dq_P}{2\pi}\chi_P(q),
\end{equation}
with $q_P=P\cdot q/M$.

The quark propagator can be decomposed into positive- and negative-energy parts as
\begin{equation}
-iJ\,S_i(Jp_i)=
\frac{\Lambda_i^+(q_\perp)}{q_P+\alpha_i M-\omega_i+i\epsilon}
+\frac{\Lambda_i^-(q_\perp)}{q_P+\alpha_i M+\omega_i-i\epsilon},
\end{equation}
where $J=+1$ for the quark and $J=-1$ for the antiquark, and
\begin{equation}
\Lambda_i^\pm(q_\perp)=\frac{1}{2\omega_i}
\left[
\frac{\slashed{P}}{M}\omega_i
\pm
\left(\slashed{q}_\perp+J m_i\right)
\right],
\qquad
\omega_i=\sqrt{m_i^2-q_\perp^2}.
\end{equation}
Here $q_\perp^2$ follows the Minkowski convention; in the meson rest frame one has
$q_\perp^2=-\mathbf{q}^2$, so that $\omega_i=\sqrt{m_i^2+\mathbf{q}^2}$.

With these definitions, the BS equation reduces to the Salpeter equation
\begin{equation}
\chi_P(q)=S_1(p_1)\eta_P(q_\perp)S_2(-p_2),
\end{equation}
with
\begin{equation}
\eta_P(q_\perp)=\int\frac{d^3k_\perp}{(2\pi)^3}
V(q_\perp,k_\perp)\varphi_P(k_\perp).
\end{equation}
Introducing the projected wave functions
\begin{equation}
\varphi_P^{\pm\pm}(q_\perp)=
\Lambda_1^\pm(q_\perp)\frac{\slashed{P}}{M}
\varphi_P(q_\perp)\frac{\slashed{P}}{M}
\Lambda_2^\pm(q_\perp),
\end{equation}
one obtains
\begin{align}
(M-\omega_1-\omega_2)\varphi_P^{++}(q_\perp)
&=\Lambda_1^+(q_\perp)\eta_P(q_\perp)\Lambda_2^+(q_\perp),\\
(M+\omega_1+\omega_2)\varphi_P^{--}(q_\perp)
&=-\Lambda_1^-(q_\perp)\eta_P(q_\perp)\Lambda_2^-(q_\perp),\\
\varphi_P^{+-}(q_\perp)&=0,\qquad
\varphi_P^{-+}(q_\perp)=0.
\end{align}

In the numerical calculation, the interaction kernel is taken in momentum
space to be of the screened Cornell type~\cite{GodfreyIsgur,EichtenQuigg},
\begin{align}
V(\mathbf{q})&=(2\pi)^3V_s(\mathbf{q})
+\gamma^0\otimes\gamma_0\,(2\pi)^3V_v(\mathbf{q}),\\
V_s(\mathbf{q})&=-\left(\frac{\lambda}{\alpha}+V_0\right)\delta^3(\mathbf{q})
+\frac{\lambda}{\pi^2}\frac{1}{(\mathbf{q}^2+\alpha^2)^2},\\
V_v(\mathbf{q})&=-\frac{2}{3\pi^2}\frac{\alpha_s(\mathbf{q})}{\mathbf{q}^2+\alpha^2}.
\end{align}
In the present calculation, $\alpha_s$ is taken to be the running coupling
\begin{equation}
\alpha_s(\mathbf{q})=\frac{12\pi}{27}\frac{1}{\log\left(a+\frac{\mathbf{q}^2}{\Lambda_{\rm QCD}^2}\right)},
\end{equation}
with $a=e=2.7183$ and $\Lambda_{\rm QCD}=0.27$ GeV.

\subsection{Relativistic Salpeter wave functions}

For the vector charmonium state with $J^{PC}=1^{--}$, the general instantaneous Salpeter wave function can be written as~\cite{Chang2005,Wang2006,Wang2007}
\begin{align}
\varphi_{1^-}(q_\perp)=&
(q_\perp\cdot \epsilon)\left[
f_1(q_\perp)
+\frac{\slashed{P}}{M}f_2(q_\perp)
+\frac{\slashed{q}_\perp}{M}f_3(q_\perp)
+\frac{\slashed{P}\slashed{q}_\perp}{M^2}f_4(q_\perp)
\right]
\nonumber\\
&+M\slashed{\epsilon}\,f_5(q_\perp)
+\slashed{\epsilon}\slashed{P}\,f_6(q_\perp)
+\left(\slashed{q}_\perp\slashed{\epsilon}
-q_\perp\cdot\epsilon\right)f_7(q_\perp)
\nonumber\\
&+\frac{1}{M}\left(
\slashed{P}\slashed{\epsilon}\slashed{q}_\perp
-\slashed{P}\,q_\perp\cdot\epsilon
\right)f_8(q_\perp),
\label{eq:salpeter_vector}
\end{align}
where $\epsilon^\mu$ is the polarization vector of the vector meson and $f_i(q_\perp)$ are scalar functions depending on $q_\perp^2$. After imposing the Salpeter constraint equations
$\varphi^{+-}=\varphi^{-+}=0$, only four of the eight scalar functions are
independent.  In the present convention, we take $f_3$, $f_4$, $f_5$, and
$f_6$ as the independent components, while $f_1$, $f_2$, $f_7$, and $f_8$
are determined by the constraint relations; the explicit relations among
them can be found in Ref.~\cite{Wang2006}.  For the
predominantly $3\,{}^3S_1$ state identified with $\psi(4040)$, the radial
wave function exhibits a characteristic multi-component nodal structure.

For the pseudoscalar heavy-light mesons with $J^P=0^-$, including $D$ and $D_s$, the general Salpeter wave function can be written as
\begin{equation}
\varphi_{0^-}(q_\perp)=
\left[
g_1(q_\perp)
+\frac{\slashed{P}}{M}g_2(q_\perp)
+\frac{\slashed{q}_\perp}{M}g_3(q_\perp)
+\frac{\slashed{P}\slashed{q}_\perp}{M^2}g_4(q_\perp)
\right]\gamma_5,
\end{equation}
where again the constraint equations reduce the number of independent scalar functions. The positive-energy wave function $\varphi^{++}_{0^-}$ is obtained by applying the positive-energy projectors and is used directly in the decay amplitude.

For the final vector heavy-light mesons with $J^P=1^-$, such as $D^*$, the Salpeter wave function has the same Lorentz structure as Eq.~\eqref{eq:salpeter_vector}, with the scalar functions $f_i$ replaced by the corresponding final-state functions $h_i$ and the polarization vector $\epsilon^\mu$ replaced by $\epsilon_f^\mu$.

In practical calculations, the positive-energy components of the Salpeter wave functions provide the dominant contribution to the strong decay amplitudes. The corresponding projected wave functions,
\begin{equation}
\varphi_P^{++}(q_\perp)=
\Lambda_1^+(q_\perp)\frac{\slashed{P}}{M}
\varphi_P(q_\perp)\frac{\slashed{P}}{M}
\Lambda_2^+(q_\perp),
\end{equation}
are obtained numerically by solving the Salpeter equation for the initial charmonium state and the final heavy-light mesons. In particular, for $\psi(4040)$ the relevant Salpeter components exhibit nodal behavior, as expected for a $3S$ state, and this multi-component nodal structure is the central dynamical ingredient of the suppressed $D\bar D$ decay amplitude.

\subsection{Relativistic \texorpdfstring{$^3P_0$}{3P0} decay amplitude}

The OZI-allowed strong decays of $\psi(4040)$ are described by the relativistic $^3P_0$ model~\cite{Micu,LeYaouanc,LeYaouancBook,Wang2013,Ding2021}, in which a light quark-antiquark pair with the vacuum quantum numbers $J^{PC}=0^{++}$ is created from the vacuum. The corresponding interaction Hamiltonian is taken as
\begin{equation}
H_I=-ig\int d^4x\,\bar{\psi}(x)\psi(x),
\end{equation}
where the coupling constant is parameterized as
\begin{equation}
g=2m_q\gamma,
\end{equation}
with $m_q$ the constituent mass of the created light quark and $\gamma$ a dimensionless strength parameter. In this work, $\gamma$ is fixed by the measured strong decay width of $\psi(3770)\to D\bar D$ and then used for the prediction of $\psi(4040)$ decays.

For the OZI-allowed process
\begin{equation}
A(P)\to B(P_1)+C(P_2),
\end{equation}
the $S$-matrix element can be written as
\begin{equation}
\langle BC|S|A\rangle=(2\pi)^4\delta^4(P-P_1-P_2)\,{\cal M}.
\end{equation}
Using the Mandelstam formalism~\cite{Mandelstam1955}, the transition amplitude is expressed as
\begin{equation}
{\cal M}
=-ig\int\frac{d^4q}{(2\pi)^4}
\mathrm{Tr}\left[
\chi_P(q)\,
S_2^{-1}(-p_2)\,
\bar{\chi}_{P_2}(q_2)\,
\bar{\chi}_{P_1}(q_1)\,
S_1^{-1}(p_1)
\right],
\end{equation}
where $q$, $q_1$, and $q_2$ are the relative momenta in the initial and final mesons, respectively. Under the instantaneous approximation and retaining the dominant positive-energy components, the amplitude is reduced to
\begin{equation}
{\cal M}
\simeq
g\int\frac{d^3q_\perp}{(2\pi)^3}
\mathrm{Tr}\left[
\frac{\slashed{P}}{M}
\varphi_P^{++}(q_\perp)
\frac{\slashed{P}}{M}
\bar{\varphi}_{P_2}^{++}(q_{2\perp})
\bar{\varphi}_{P_1}^{++}(q_{1\perp})
\right].
\label{eq:amp_reduced}
\end{equation}
This approximation is standard in relativistic Salpeter applications and captures the dominant contribution, although subleading components may still affect the quantitative value of amplitudes that are numerically close to zero.
The corresponding partial decay width is
\begin{equation}
\Gamma(A\to BC)
=
\frac{|\mathbf{P}_1|}{8\pi M_A^2}
\frac{1}{2J_A+1}
\sum_{\lambda}
|{\cal M}|^2,
\label{eq:width}
\end{equation}
where $J_A$ is the spin of the initial state and $|\mathbf{P}_1|$ is the three-momentum of the final mesons in the rest frame of the initial meson,
\begin{equation}
|\mathbf{P}_1|
=
\frac{\sqrt{[M_A^2-(M_B+M_C)^2][M_A^2-(M_B-M_C)^2]}}{2M_A}.
\end{equation}

In the present work, we focus on the channels
\begin{equation}
\psi(4040)\to D\bar D,\qquad
\psi(4040)\to D\bar D^*+{\rm c.c.},\qquad
\psi(4040)\to D_s\bar D_s.
\end{equation}
For later discussion of the nodal mechanism, it is useful to introduce
the momentum-dependent overlap integrand after the angular integration.
For the pseudoscalar-pseudoscalar final states,
\[
D^0\bar D^0,\qquad D^+D^-,\qquad D_s\bar D_s,
\]
the decay amplitude can be written schematically as
\begin{equation}
{\cal M}_{i}
=
g\int_0^\infty dq\, J_i(q),
\label{eq:Mi_PP}
\end{equation}
where
\begin{equation}
J_i(q)
=
\int_0^\pi d\theta\,
\frac{q^2\sin\theta}{(2\pi)^3}\,
{\cal T}_i(q,\theta).
\label{eq:J_PP}
\end{equation}
Here ${\cal T}_i(q,\theta)$ denotes the full trace structure obtained
from the relativistic $^3P_0$ transition amplitude, including all
relevant Salpeter components, kinematic factors, and final-state wave
functions. Therefore $J_i(q)$ should not be identified with a single
radial component of the initial wave function, but with the complete
channel-dependent overlap integrand.

For the pseudoscalar-vector final states,
\[
D^0\bar D^{*0},\qquad D^+D^{*-},
\]
the relativistic amplitude contains two independent structures, denoted
by ${\cal M}_{i,12}$ and ${\cal M}_{i,21}$. We define
\begin{equation}
{\cal M}_{i,a}
=
g\int_0^\infty dq\, J_{i,a}(q),
\qquad
a=12,21,
\label{eq:Mi_PV}
\end{equation}
with
\begin{equation}
J_{i,a}(q)
=
\int_0^\pi d\theta\,
\frac{q^2\sin\theta}{(2\pi)^3}\,
{\cal T}_{i,a}(q,\theta),
\qquad
a=12,21 .
\label{eq:J_PV}
\end{equation}
The two structures are integrated separately and enter the partial width
incoherently,
\begin{equation}
\Gamma_i(PV)
\propto
|{\cal M}_{i,12}|^2+|{\cal M}_{i,21}|^2 ,
\label{eq:PV_incoherent}
\end{equation}
after the spin average and phase-space factors, where the interference term between the two structures vanishes. Thus they should not be
added at the integrand level.

\subsection{Numerical strategy}

The masses and BS wave functions of the initial and final mesons are obtained by solving the Salpeter equation numerically. For the vector $3\,{}^3S_1$ charmonium state, the BS equation gives a mass of $4051$ MeV, and the corresponding wave function is used as the reference $3S$ state in the decay calculation. The decay parameter $\gamma$ is determined independently from $\psi(3770)\to D\bar D$ and then kept fixed in the calculation of $\psi(4040)$ decays. For the central prediction, the phase space is evaluated with the experimental mass $M_A=4040$ MeV, while the overlap amplitudes are computed with the BS wave function of the nearby theoretical $3S$ eigenstate at $4051$ MeV; results at $M_A=4051$ MeV are also presented for comparison. In this sense, the $4051$ MeV state is the self-consistent BS eigenstate used to define the initial wave function, whereas $4040$ MeV is the experimental pole mass used for the physical phase space.

The $V_0$ parameter of the screened Cornell potential is fitted globally to reproduce the charmonium spectrum, including $J/\psi$ and $\psi(2S)$; it cannot be adjusted for $\psi(4040)$ alone without degrading the description of the lower states. The resulting $11$ MeV mismatch between the BS eigenmass ($4051$ MeV) and the experimental pole ($4040$ MeV) is therefore an inherent feature of the present potential model rather than a tunable parameter. Its main effect is quantitative rather than qualitative: because the physical region lies near a nodal zero, the absolute value of the $D\bar D$ width is sensitive to this shift, whereas the suppression hierarchy, the dip structure, and the channel selectivity remain stable across the scanned mass range. In the mass scan, the same wave function is kept fixed while $M_A$ is varied, so that the intrinsic sign-changing structure of the overlap amplitude can be separated from the purely kinematic effect of moving the initial mass through the nodal region. The fixed-wave-function use of the BS eigenstate together with nearby physical masses should therefore be understood as a diagnostic setup for isolating the nodal mechanism, rather than as a fully self-consistent treatment of mass shifts.

\section{Inputs and wave functions}

\subsection{Model parameters}

In this section, we specify the model inputs used in the numerical calculation and summarize the properties of the resulting bound-state wave functions. The same instantaneous Bethe--Salpeter framework is employed for the initial charmonium state and the final heavy-light mesons, so that the decay amplitudes can be evaluated consistently.

The constituent quark masses, the parameters in the screened Cornell kernel, and the decay strength parameter $\gamma$ are the main inputs of the present calculation. The charmonium and heavy-light meson wave functions are obtained by solving the Salpeter equation with a common set of model parameters. The quark masses and potential parameters are chosen to reproduce the relevant meson spectra~\cite{GodfreyIsgur,EichtenQuigg,Ding2021}, whereas the $^3P_0$ strength parameter is fixed independently from the measured decay width of $\psi(3770)\to D\bar D$.
In the present setup, the quark masses, the parameters $\lambda$ and $\alpha$, and the running-coupling inputs $a$ and $\Lambda_{\rm QCD}$ are taken as common inputs, while the constant term $V_0$ is allowed to vary between meson sectors as an effective offset in the standard Salpeter phenomenology.

For clarity, the main numerical inputs used in this work are collected in
Tables~\ref{tab:parameters} and~\ref{tab:V0values}.  Table~\ref{tab:parameters} lists
the constituent quark masses and the common potential parameters, while
Table~\ref{tab:V0values} gives the channel-dependent values of the constant
term $V_0$ used for the relevant meson systems.

\begin{table}[t]
\caption{Input parameters used in the Bethe--Salpeter equation and the relativistic $^3P_0$ model.}
\label{tab:parameters}
\begin{ruledtabular}
\begin{tabular}{cc}
Parameter & Value \\
\hline
$m_c$ & $1.620$ GeV \\
$m_u$ & $0.305$ GeV \\
$m_d$ & $0.311$ GeV \\
$m_s$ & $0.500$ GeV \\
$\lambda$ & $0.210~\mathrm{GeV}^2$ \\
$\alpha$ & $0.06$ \\
$a=e$ & $2.7183$ \\
$\Lambda_{\rm QCD}$ & $0.27$ GeV \\
$\gamma$ & $0.526\pm0.023$, fitted from $\psi(3770)\to D^0\bar D^0$ and $D^+D^-$ \\
\end{tabular}
\end{ruledtabular}
\end{table}

\begin{table}[t]
\caption{Values of the constant term $V_0$ used for the meson systems relevant to the present calculation.}
\label{tab:V0values}
\begin{ruledtabular}
\begin{tabular}{ccc}
System & $J^P$ & $V_0$ (GeV) \\
\hline
$c\bar c$ & $1^-$ & $-0.1756$ \\
$c\bar u$ & $0^-$ & $-0.375$ \\
$c\bar d$ & $0^-$ & $-0.375$ \\
$c\bar s$ & $0^-$ & $-0.432$ \\
$c\bar u$ & $1^-$ & $-0.110$ \\
$c\bar d$ & $1^-$ & $-0.110$ \\
\end{tabular}
\end{ruledtabular}
\end{table}

\subsection{Determination of the \texorpdfstring{$^3P_0$}{3P0} strength from \texorpdfstring{$\psi(3770)\to D\bar D$}{psi(3770) to DD}}

A crucial input of the present calculation is the pair-creation strength parameter $\gamma$ in the relativistic $^3P_0$ model. Rather than fitting $\gamma$ directly to the decay data of $\psi(4040)$, we determine it independently from the measured width of the nearby open-charm process $\psi(3770)\to D\bar D$~\cite{PDG,Ablikim2004,Ablikim2006,Besson2006}. Since $\psi(3770)$ lies close to $\psi(4040)$ in the charmonium spectrum and its dominant strong decay mode is experimentally well established, it provides a natural calibration channel for the pair-creation strength.

More specifically, we calculate the decay width of $\psi(3770)\to D\bar D$ within the same Bethe--Salpeter plus relativistic $^3P_0$ framework and fix $\gamma$ by requiring the theoretical result to reproduce the experimental central value. The BS calculation gives the $1\,{}^3D_1$ mass at $3779$ MeV, only $5$ MeV above the experimental $\psi(3770)$ mass, indicating that a pure $D$-wave assignment appears adequate for the present calibration purpose; possible $S$-$D$ mixing effects are therefore expected to be small and are not included in the present determination of $\gamma$. The resulting value, $\gamma=0.526\pm0.023$, is consistent with typical values used in relativistic $^3P_0$ analyses. Once $\gamma$ is fixed in this way, no further adjustment is made in the calculation of the open-charm decays of $\psi(4040)$. The resulting hierarchy for the $\psi(4040)$ decays considered here should therefore be viewed as a consequence of the relativistic wave functions and the channel-dependent overlap integrals, rather than as the result of a direct fit to the $\psi(4040)$ data.

In practice, the uncertainty of the input width of $\psi(3770)\to D\bar D$ induces a corresponding uncertainty in $\gamma$, which is then propagated to the predicted partial widths of $\psi(4040)$. We estimate this effect by varying $\gamma$ within the range allowed by the calibration channel. This procedure not only reduces the model arbitrariness, but also makes the predictive power of the present analysis more transparent.
The quoted uncertainty of $\gamma$ is obtained by propagating the experimental uncertainty of the calibration width.

\subsection{Mass spectrum}

Before discussing the strong decays, one should verify that the adopted parameter set gives a reasonable description of the relevant meson masses. In Table~\ref{tab:spectrum}, we list the calculated masses of the low-lying charmonium states and the final-state open-charm mesons, together with the corresponding experimental values. In particular, the mass of the vector $3\,{}^3S_1$ charmonium state should be close to the $\psi(4040)$ region, while the masses of $D$, $D^*$, and $D_s$ enter directly into the phase space of the decay channels under consideration.

\begin{table*}[t]
\caption{Calculated masses of relevant charmonium and open-charm mesons, compared with experimental values.}
\label{tab:spectrum}
\begin{ruledtabular}
\begin{tabular}{cccc}
State & Quark model assignment & This work (MeV) & Experiment (MeV) \\
\hline
$J/\psi$ & $1\,{}^3S_1$ & $3096$ & $3096$ \\
$\psi(2S)$ & $2\,{}^3S_1$ & $3689$ & $3686$ \\
$\psi(3770)$ & $1\,{}^3D_1$ & $3779$ & $3774$ \\
$\psi(4040)$ & $3\,{}^3S_1$ & $4051$ & $4040$ \\
$D^0$ & $0^-$ & $1865$ & $1865$ \\
$D^+$ & $0^-$ & $1870$ & $1870$ \\
$D^{*0}$ & $1^-$ & $2007$ & $2007$ \\
$D^{*+}$ & $1^-$ & $2010$ & $2010$ \\
$D_s$ & $0^-$ & $1968$ & $1968$ \\
\end{tabular}
\end{ruledtabular}
\end{table*}

A reasonable reproduction of the spectrum is an important prerequisite for the decay calculation, especially for $\psi(4040)$, whose decay amplitudes are found to be highly sensitive to the available phase space and to the momentum region probed by the radial wave function. In the present calculation, the $3\,{}^3S_1$ charmonium mass is obtained as $4051$ MeV, which is slightly higher than the experimental mass of $\psi(4040)$ but still sufficiently close to support the conventional assignment. This small difference is physically important in the present problem, because the $D\bar D$ width is found to vary rapidly with the initial mass when the amplitude is evaluated near a nodal zero.

\subsection{Wave functions and node structure}

\begin{figure}[t]
\centering
\includegraphics[width=0.65\textwidth]{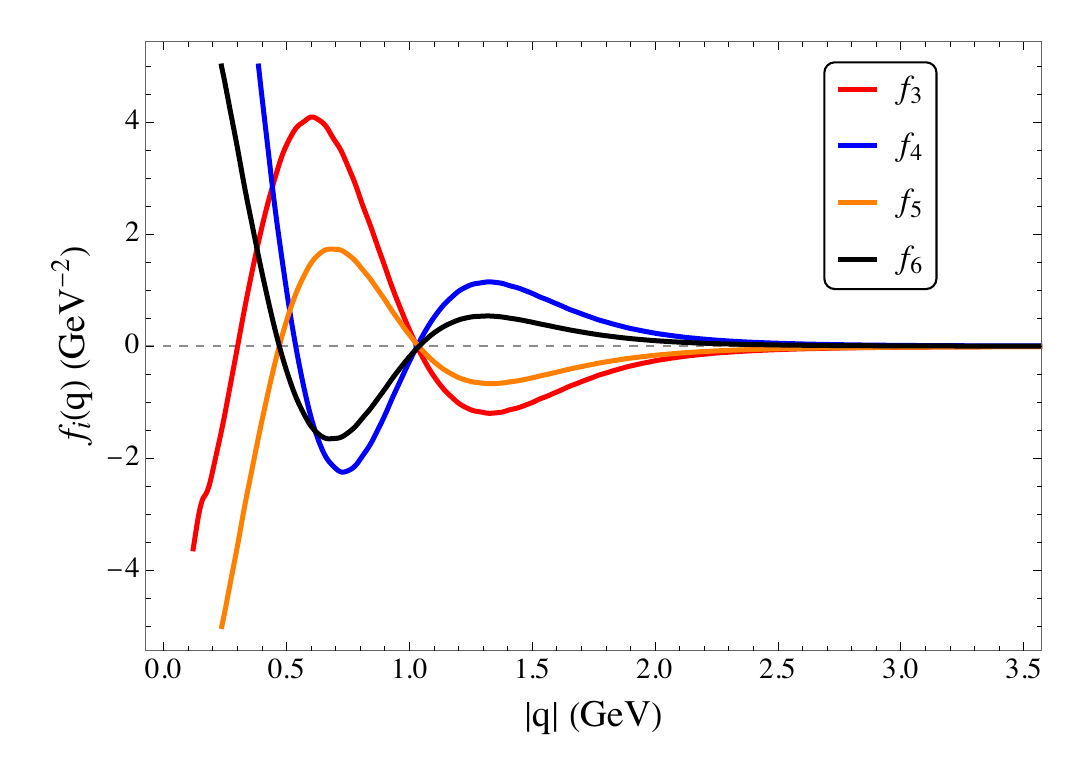}
\caption{Radial components of the positive-energy Salpeter wave function for the $3\,{}^3S_1$ charmonium state identified with $\psi(4040)$. The curves correspond to the independent components $f_3$, $f_4$, $f_5$, and $f_6$. All components exhibit nodal behavior, while their node positions are different, reflecting the relativistic multi-component structure of the Bethe--Salpeter wave function.}
\label{fig:wavefunction}
\end{figure}

The central physical ingredient of the present work is the component-dependent nodal structure of the $\psi(4040)$ wave function. The radial components $f_3$, $f_4$, $f_5$, and $f_6$ of the positive-energy Salpeter wave function are shown in Fig.~\ref{fig:wavefunction}. All these components exhibit nodal behavior, as expected for a radially excited $3S$ state. However, the node positions are not identical for different components. This reflects the relativistic multi-component structure of the Bethe--Salpeter wave function and shows that the nodal structure cannot be characterized by a single zero of the full wave function.

Consequently, the decay suppression should not be attributed to the vanishing of an individual radial component. Instead, the relevant quantity is the sign structure of the full overlap integrand entering the decay amplitude.

\subsection{Expected sensitivity of the decay amplitudes}

The decay amplitudes of $\psi(4040)$ can be expressed schematically as overlap integrals of the initial and final wave functions. Because the $\psi(4040)$ radial wave function contains a node, the sign structure of the integrand becomes crucial. If the momentum region favored by a given decay channel overlaps strongly with the nodal region, the integral is suppressed by cancellation. If not, the corresponding channel remains sizable.

This observation indicates that the $D\bar D$ mode is abnormally small, while the $D\bar D^*$ and $D_s\bar D_s$ channels remain much larger. A quantitative demonstration of this mechanism is given in the next section through the explicit evaluation of the partial widths and the channel-dependent overlap integrals.

\section{Open-charm decays of \texorpdfstring{$\psi(4040)$}{psi(4040)}}

In this section, we present the numerical results for the open-charm strong decays of $\psi(4040)$ and identify the mechanism responsible for the observed hierarchy of decay channels. Throughout this section, $\psi(4040)$ is treated as the conventional $3\,{}^3S_1$ charmonium state. The relativistic BS wave function used in the amplitudes is the nearby $3S$ eigenstate obtained at $4051$ MeV, while the quoted central widths use the physical phase space at $M_A=4040$ MeV. As emphasized in Sec.~II, this fixed-wave-function setup is used to expose the dynamical sensitivity of the channels to the nodal structure rather than to construct a fully coupled-channel description of the mass shift itself.

\subsection{Baseline decay widths}

We first evaluate the partial widths of the channels
\begin{equation}
\psi(4040)\to D\bar D,\qquad
\psi(4040)\to D\bar D^*+{\rm c.c.},\qquad
\psi(4040)\to D_s\bar D_s.
\end{equation}
The numerical results display a clear hierarchy among these modes,
\begin{equation}
\Gamma(D\bar D^*+{\rm c.c.}) \gg \Gamma(D_s\bar D_s) \gg \Gamma(D\bar D),
\end{equation}
in agreement with the qualitative experimental pattern~\cite{PDG}. The striking feature is the extreme suppression of $D\bar D$: although the channel has ample phase space, its width is smaller than those of $D\bar D^*$ and $D_s\bar D_s$ by orders of magnitude. The scope of the present calculation is deliberately limited to the three channels $D\bar D$, $D\bar D^*+{\rm c.c.}$, and $D_s\bar D_s$, which most directly expose the puzzle of the suppressed $D\bar D$ mode. The $D^*\bar D^*$ channel, though experimentally important, is a near-threshold $P$-wave final state whose quantitative treatment requires a dedicated analysis of coupled-channel and phase-space effects beyond the scope of this work. The present results are therefore not intended as a complete description of the $\psi(4040)$ total width, but as an identification of the amplitude-level origin of the $D\bar D$ suppression within a conventional $3\,{}^3S_1$ picture.

Numerically, for the central choice $M_A=4040$ MeV, we obtain
\begin{align}
\Gamma(D^0\bar D^0) &=(0.0101\pm0.0009)~\text{MeV},\label{eq:width_D0D0_4040}\\
\Gamma(D^+D^-) &=(0.0182\pm0.0016)~\text{MeV},\label{eq:width_DpDm_4040}\\
\Gamma(D\bar D) &=(0.0283\pm0.0025)~\text{MeV},\label{eq:width_DD_4040}\\
\Gamma(D^0\bar D^{*0}+{\rm c.c.}) &=(10.6636\pm0.9308)~\text{MeV},\label{eq:width_D0Dstar0_4040}\\
\Gamma(D^+D^{*-}+{\rm c.c.}) &=(16.1850\pm1.4128)~\text{MeV},\label{eq:width_DpDstarm_4040}\\
\Gamma(D\bar D^*+{\rm c.c.}) &=(26.8486\pm2.3436)~\text{MeV},\label{eq:width_DDstar_4040}\\
\Gamma(D_s\bar D_s) &=(7.9110\pm0.4790)~\text{MeV}.\label{eq:width_DsDs_4040}
\end{align}
For clarity, the grouped partial widths at $M_A=4040$ MeV are collected
in Table~\ref{tab:widths4040}. The charged and neutral contributions are
given explicitly in Eqs.~\eqref{eq:width_D0D0_4040}--\eqref{eq:width_DsDs_4040}, while Table~\ref{tab:widths4040}
summarizes both the grouped widths at $M_A=4040$ MeV and the qualitative
comparison with experiment.

To make the role of the $11$ MeV mass mismatch explicit, Table~\ref{tab:masscompare4040} also lists the results obtained at $M_A=4051$ MeV with the same BS wave function. This provides a minimal robustness check within the fixed-wave-function setup: the hierarchy and the ordering of the sizable channels remain stable, whereas the nearly vanishing $D\bar D$ channel is exceptionally sensitive to the mass shift because it lies near a nodal zero. For the $D\bar D^*$ modes, the amplitudes are first evaluated for the explicitly calculated charge channels, $D^0\bar D^{*0}$ and $D^+D^{*-}$. The charge-conjugated channels have the same widths in the present calculation and are included through a multiplicity factor of two. The quantities quoted below with ``$+{\rm c.c.}$'' already include this factor.

At $M_A=4051$ MeV, the corresponding widths become
\begin{align}
\Gamma(D^0\bar D^0) &=(0.1208\pm0.0105)~\text{MeV},\\
\Gamma(D^+D^-) &=(0.0141\pm0.0012)~\text{MeV},\\
\Gamma(D\bar D) &=(0.1349\pm0.0117)~\text{MeV},\\
\Gamma(D^0\bar D^{*0}+{\rm c.c.}) &=(8.6598\pm0.7558)~\text{MeV},\\
\Gamma(D^+D^{*-}+{\rm c.c.}) &=(13.5112\pm1.1810)~\text{MeV},\\
\Gamma(D\bar D^*+{\rm c.c.}) &=(22.1710\pm1.9368)~\text{MeV},\\
\Gamma(D_s\bar D_s) &=(7.8490\pm0.4753)~\text{MeV}.
\end{align}

\begin{table}[t]
\caption{Grouped partial widths of the open-charm decay channels of $\psi(4040)$ at $M_A=4040$ MeV, together with the corresponding qualitative comparison with experiment. The experimental information is included only at the qualitative level, since quantitative branching ratios for the individual open-charm channels are not well established.}
\label{tab:widths4040}
\begin{ruledtabular}
\begin{tabular}{ccc}
Channel & Experiment~\cite{PDG} & $\Gamma$ (MeV) \\
\hline
$D\bar D$ & Strongly suppressed & $(0.0283\pm0.0025)$ (suppressed) \\
$D\bar D^*+{\rm c.c.}$ & Dominant & $(26.8486\pm2.3436)$ (dominant) \\
$D_s\bar D_s$ & Sizable & $(7.9110\pm0.4790)$ (sizable) \\
\end{tabular}
\end{ruledtabular}
\end{table}

The strong suppression of the $D\bar D$ amplitude can be understood, as shown below, as a consequence of the component-dependent nodal structure of the $3S$ Bethe--Salpeter wave function.

\subsection{Comparison with experiment and other theoretical calculations}

Although quantitative branching ratios for the individual open-charm channels of $\psi(4040)$ are not well established experimentally, the qualitative hierarchy is firmly supported by the available data~\cite{PDG}: $D\bar D^*+{\rm c.c.}$ is the dominant mode, $D_s\bar D_s$ is sizable, and $D\bar D$ is strongly suppressed. The present calculation reproduces this qualitative pattern, as summarized in Table~\ref{tab:widths4040}.

This hierarchy has also been obtained in previous theoretical studies. In particular, Barnes, Godfrey, and Swanson~\cite{Barnes2005} used the nonrelativistic $^3P_0$ model with a harmonic-oscillator wave function and found that the $3S$ assignment for $\psi(4040)$ gives a strongly suppressed $D\bar D$ width due to nodal effects, while $D\bar D^*$ and $D_s\bar D_s$ remain sizable. Eichten and Quigg~\cite{Eichten2019} employed the Cornell potential model and also confirmed the qualitative hierarchy $\Gamma(D\bar D^*)\gg\Gamma(D_s\bar D_s)\gg\Gamma(D\bar D)$ for the $3\,{}^3S_1$ assignment. The present results are consistent with these earlier findings at the qualitative level, while making the cancellation mechanism more explicit at the amplitude level. The new contribution of this work is not the reproduction of the hierarchy itself, but the explicit identification of its amplitude-level origin: the channel-dependent overlap integrand $J_i(q)$ undergoes sign-changing cancellation specifically in the $D\bar D$ channel, while $D\bar D^*$ and $D_s\bar D_s$ avoid this severe cancellation because they probe different momentum regions of the same initial-state wave function.

\subsection{Nodal filtering and the suppressed \texorpdfstring{$D\bar D$}{DD} channel}

The nodal mechanism becomes explicit once one examines the
momentum-dependent overlap integrands defined in
Eqs.~\eqref{eq:J_PP} and \eqref{eq:J_PV}. For the
pseudoscalar-pseudoscalar channels, the relevant quantity is $J_i(q)$.
For the pseudoscalar-vector channels, the two independent integrands
$J_{i,12}(q)$ and $J_{i,21}(q)$ must be kept separate because the
partial width is proportional to
$|{\cal M}_{i,12}|^2+|{\cal M}_{i,21}|^2$, rather than to
$|{\cal M}_{i,12}+{\cal M}_{i,21}|^2$.

For the pseudoscalar-pseudoscalar channels we define the cumulative
integral
\begin{equation}
I_i(q_{\max})
=
\int_0^{q_{\max}} dq\,J_i(q),
\qquad
i=D^0\bar D^0,\ D^+D^-,\ D_s\bar D_s .
\label{eq:I_PP}
\end{equation}
For the pseudoscalar-vector channels we define
\begin{equation}
I_{i,a}(q_{\max})
=
\int_0^{q_{\max}} dq\,J_{i,a}(q),
\qquad
a=12,21 .
\label{eq:I_PV}
\end{equation}
In the limit of large $q_{\max}$, the unnormalized cumulative integrals
approach the corresponding decay amplitudes up to the overall factor
$g$.

\begin{figure}[t]
\centering
\includegraphics[width=0.95\textwidth]{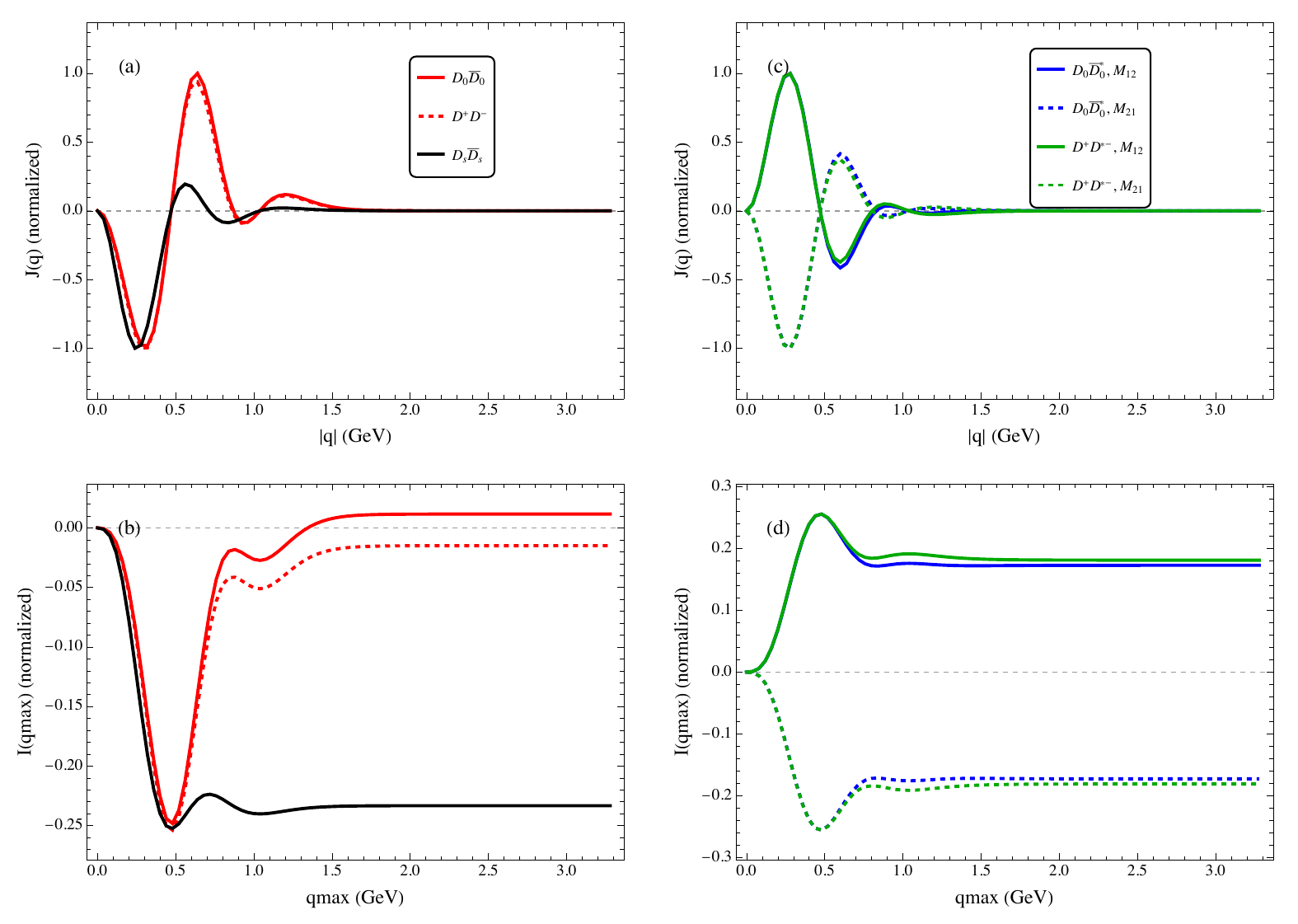}
\caption{
Normalized momentum-dependent overlap integrands and cumulative
integrals for the open-charm decay channels of $\psi(4040)$.
Panels (a) and (b) show the pseudoscalar-pseudoscalar channels
$D^0\bar D^0$, $D^+D^-$, and $D_s\bar D_s$, while panels (c) and (d)
show the pseudoscalar-vector channels $D^0\bar D^{*0}$ and
$D^+D^{*-}$. For the $D\bar D^*$ final states, the two independent
amplitude structures ${\cal M}_{12}$ and ${\cal M}_{21}$ are displayed
separately, since they enter the partial width incoherently as
$|{\cal M}_{12}|^2+|{\cal M}_{21}|^2$. Each integrand is normalized by
its maximum absolute value in order to emphasize the sign-changing
structure. The $D^0\bar D^0$ and $D^+D^-$ cumulative integrals are
driven back toward very small final values, indicating strong
destructive cancellation in the decay amplitudes. In contrast, the
$D_s\bar D_s$ and $D\bar D^*$ channels retain sizable cumulative
contributions and therefore do not undergo the same severe cancellation.
The charge-conjugated $D\bar D^*$ modes are not shown separately,
because they have the same cancellation pattern and are included by the
corresponding multiplicity factor in the partial widths.
}
\label{fig:overlap_integrands}
\end{figure}

In Fig.~\ref{fig:overlap_integrands}, we show the normalized overlap
integrands and the corresponding cumulative integrals. The normalization
is introduced only for visualization,
\begin{equation}
\widehat J_i(q)
=
\frac{J_i(q)}{\max_q |J_i(q)|},
\qquad
\widehat J_{i,a}(q)
=
\frac{J_{i,a}(q)}{\max_q |J_{i,a}(q)|}.
\label{eq:J_normalized}
\end{equation}
The cumulative curves shown in Fig.~\ref{fig:overlap_integrands} are
obtained by integrating $\widehat J_i(q)$ and $\widehat J_{i,a}(q)$. Therefore, the
figure is intended to display the sign-changing and cancellation
patterns, while the absolute magnitudes of the partial widths are given
in Tables~\ref{tab:widths4040} and \ref{tab:masscompare4040}.

The left panels of Fig.~\ref{fig:overlap_integrands} show the
pseudoscalar-pseudoscalar channels. For $D^0\bar D^0$ and $D^+D^-$,
the integrands receive comparable positive and negative contributions
from different momentum regions, so the cumulative integrals are driven
back toward nearly zero after passing through the sign-changing region.
This is the suppression mechanism in its most direct form: the physical
$D\bar D$ amplitude is small because the full relativistic overlap
integral nearly cancels.

The contrast with $D_s\bar D_s$ is equally important. Its cumulative
integral remains clearly away from zero, showing that this channel probes
a different momentum region of the same initial $3S$ wave function and
therefore avoids the severe cancellation that suppresses $D\bar D$.
The right panels show the pseudoscalar-vector channels. The two
independent structures ${\cal M}_{12}$ and ${\cal M}_{21}$ are displayed
separately for $D^0\bar D^{*0}$ and $D^+D^{*-}$. Even when these
structures have opposite signs, they contribute to the width
incoherently through $|{\cal M}_{12}|^2+|{\cal M}_{21}|^2$, so there is
no analogous cancellation in the physical $D\bar D^*$ rate. The sizable
final values of their cumulative integrals therefore explain why the
$D\bar D^*$ modes remain large.

Figure~\ref{fig:overlap_integrands} shows that the decay amplitude
acts as a momentum-space filter: channels whose overlap integrals sample
the sign-changing region are strongly suppressed, whereas channels that
sample neighboring regions remain sizable. In the BS framework, this is
the amplitude-level content of the nodal mechanism underlying the
$\psi(4040)$ hierarchy.

\subsection{Mass dependence and the dip structure}

\begin{figure}[t]
\centering
\includegraphics[width=1\textwidth]{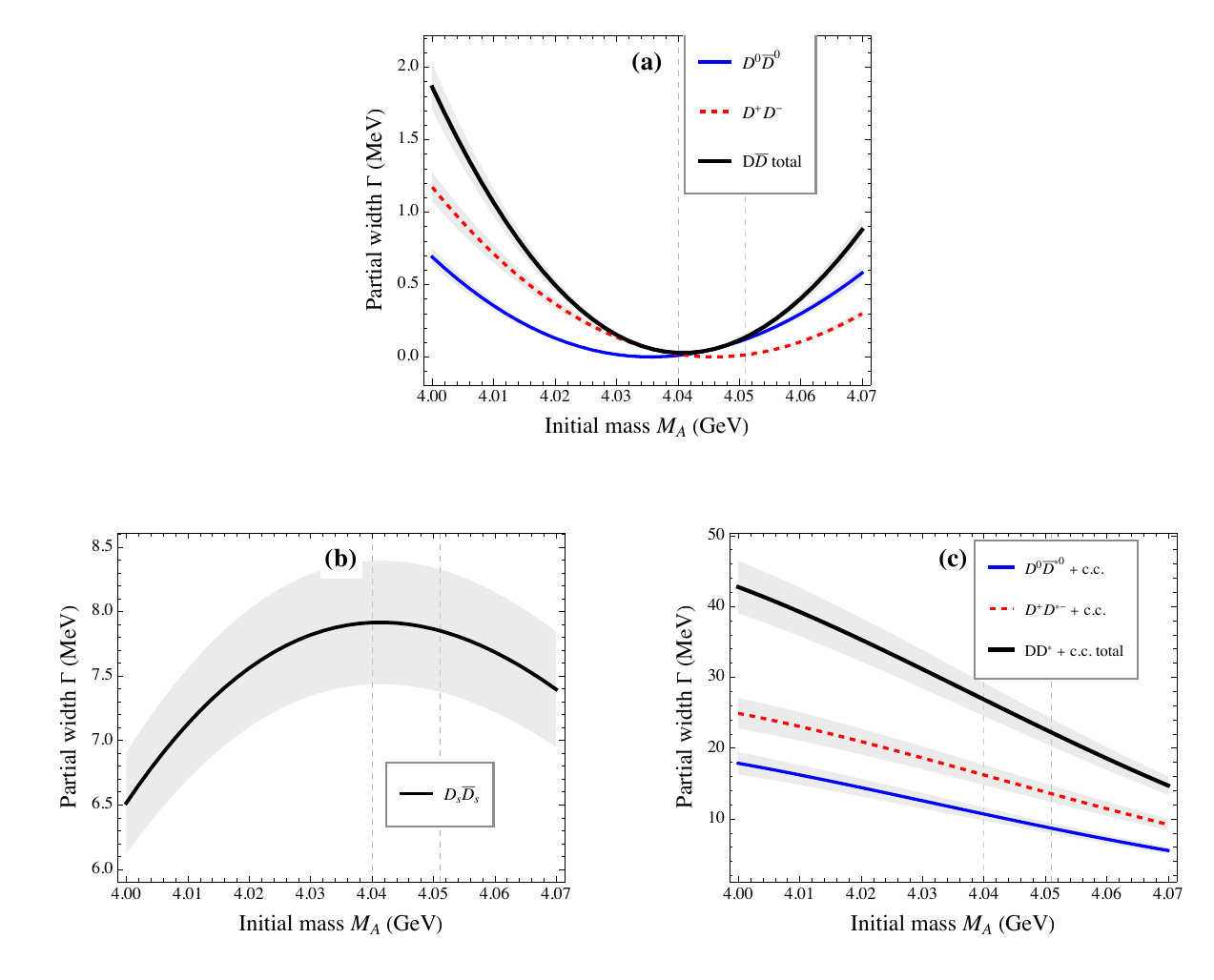}
\caption{Partial widths of $\psi(4040)$ as functions of the initial mass $M_A$.
Panel (a) shows the $D\bar D$ sector, including
$D^0\bar D^0$, $D^+D^-$, and their sum.
A pronounced dip appears in the $D\bar D$ width, indicating that the
physical region lies close to a zero of the channel-dependent overlap
amplitude.
Panel (b) shows the $D_s\bar D_s$ channel, which remains at the
several-MeV level and does not exhibit a near-zero dip.
Panel (c) shows the $D\bar D^*+{\rm c.c.}$ sector, including the neutral,
charged, and total contributions.  These widths vary smoothly and remain
sizable, confirming that the strong suppression is specific to the
$D\bar D$ channel.  The shaded bands indicate the uncertainty propagated
from the fitted pair-creation strength $\gamma$, and the vertical dashed
lines mark $M_A=4040$ MeV and $M_A=4051$ MeV.}
\label{fig:massscan4040}
\end{figure}

If the suppression of the $D\bar D$ channel is caused by a near
cancellation around the nodal region, the corresponding partial width
should be highly sensitive to the initial mass. To test this expectation,
we vary the mass $M_A$ of the initial charmonium state in the range
$4.00$--$4.07$ GeV while keeping the same Bethe--Salpeter wave function
for the nearby $3\,{}^3S_1$ eigenstate. This scan is the main check that
the central $4040/4051$ setup is not being used as an arbitrary tuning:
it shows explicitly how the physical phase space moves a fixed nearby $3S$
wave function across the sign-changing region of the overlap amplitude.
The resulting mass dependence of the partial widths is shown in
Fig.~\ref{fig:massscan4040}.

The $D\bar D$ sector, shown in Fig.~\ref{fig:massscan4040}(a), exhibits a
pronounced dip structure.  Both the neutral and charged channels are
driven close to zero in the vicinity of the physical $\psi(4040)$ mass,
although their minima occur at slightly different positions because of
the charged-neutral mass splitting.  This behavior is a direct signature
of a node-induced cancellation in the channel-dependent overlap amplitude.

By contrast, the $D_s\bar D_s$ channel in Fig.~\ref{fig:massscan4040}(b)
remains at the several-MeV level and changes smoothly with $M_A$.
This provides an important comparison with the $D\bar D$ channel:
although both are pseudoscalar-pseudoscalar final states, the
$D_s\bar D_s$ channel does not probe the same destructive-cancellation
region of the overlap integral.  The smallness of the $D\bar D$ width is
therefore not a generic feature of all $0^-0^-$ channels, but a
channel-dependent nodal filtering effect.

The $D\bar D^*+{\rm c.c.}$ widths, displayed in
Fig.~\ref{fig:massscan4040}(c), also remain sizable and do not show a
near-zero dip.  They decrease smoothly in the scanned mass range.  This
does not contradict the increase of the available phase space with
$M_A$, because the partial width is controlled by both the phase-space
factor and the dynamical overlap amplitudes.  In the present case, the
moderate phase-space enhancement is overcompensated by the decrease of
the overlap contribution
$|{\cal M}_{12}|^2+|{\cal M}_{21}|^2$, leading to a decreasing width.
Thus the mass dependence again reflects the channel-dependent momentum
filtering of the relativistic decay amplitude.

Indeed, when the initial mass is shifted from $4040$ MeV to $4051$ MeV,
the neutral and charged $D\bar D$ channels change in a markedly
nonuniform way, while the $D\bar D^*$ and $D_s\bar D_s$ modes remain
of the same order of magnitude.

For comparison, Table~\ref{tab:masscompare4040} lists the partial widths
obtained at both the experimental mass $M_A=4040$ MeV and the BS eigenmass
$M_A=4051$ MeV associated with the reference $3S$ wave function.
This comparison illustrates that the $D\bar D$ channel is exceptionally
unstable under small shifts of the initial mass, whereas the hierarchy
among the nonsuppressed channels remains qualitatively unchanged.

\begin{table}[t]
\caption{Comparison of the partial widths of $\psi(4040)$ at two representative initial masses.}
\label{tab:masscompare4040}
\begin{ruledtabular}
\begin{tabular}{ccc}
Channel & $M_A=4040$ MeV: $\Gamma$ (MeV) & $M_A=4051$ MeV: $\Gamma$ (MeV) \\
\hline
$D^0\bar D^0$ & $(0.0101\pm0.0009)$ & $(0.1208\pm0.0105)$ \\
$D^+D^-$ & $(0.0182\pm0.0016)$ & $(0.0141\pm0.0012)$ \\
$D\bar D$ & $(0.0283\pm0.0025)$ & $(0.1349\pm0.0117)$ \\
$D^0\bar D^{*0}+{\rm c.c.}$ & $(10.6636\pm0.9308)$ & $(8.6598\pm0.7558)$ \\
$D^+D^{*-}+{\rm c.c.}$ & $(16.1850\pm1.4128)$ & $(13.5112\pm1.1810)$ \\
$D\bar D^*+{\rm c.c.}$ & $(26.8486\pm2.3436)$ & $(22.1710\pm1.9368)$ \\
$D_s\bar D_s$ & $(7.9110\pm0.4790)$ & $(7.8490\pm0.4753)$ \\
\end{tabular}
\end{ruledtabular}
\end{table}

The dip structure of $\Gamma(D\bar D)$ as a function of $M_A$ therefore
provides a particularly direct piece of evidence for the nodal mechanism.
It shows that the suppression is not caused by a small overall coupling,
but by the fact that the physical region lies close to a zero of the
channel-dependent overlap integral. The absence of a comparable dip in
$D_s\bar D_s$ and $D\bar D^*+{\rm c.c.}$ demonstrates that this is a
specific dynamical cancellation in $D\bar D$, not a generic artifact of
the common BS wave function or of the fixed-wave-function setup. In
other words, the same fixed nearby $3S$ wave function produces both a
stable hierarchy among the sizable channels and a highly unstable
near-zero $D\bar D$ channel, which is exactly the pattern expected from a
channel-selective nodal zero.

\subsection{Isospin-limit test and charged-neutral channels}

Because the $D\bar D$ amplitude is strongly suppressed by nodal
cancellation, even a small difference between the charged and neutral
$D$-meson masses can produce a visible relative difference between the
corresponding partial widths. At first sight, such a difference might be
misinterpreted as a signal of strong isospin breaking. In the present
case, however, it simply reflects the amplification of small
isospin-breaking inputs in the final-state masses and wave functions when
the $D\bar D$ amplitude is already close to a nodal zero.

To demonstrate this point explicitly, we first calculate the physical
charged and neutral channels separately,
\begin{equation}
\psi(4040)\to D^0\bar D^0,\qquad
\psi(4040)\to D^+D^- .
\end{equation}
We then repeat the calculation in the isospin limit.  In this test, a
common light-quark mass is used,
\begin{equation}
m_u=m_d\equiv m_q^{\rm iso}=0.308~{\rm GeV},
\end{equation}
and the corresponding isospin-limit $D$-meson wave function is recalculated.
The final-state masses are also taken to be degenerate,
\begin{equation}
M_{D^0}=M_{D^+}\equiv M_D^{\rm iso}.
\end{equation}
The resulting width therefore corresponds to each charge channel in the
isospin limit.  The corresponding results are summarized in
Table~\ref{tab:isospin4040}.

\begin{table}[t]
\caption{
Charged and neutral $D\bar D$ partial widths of $\psi(4040)$, together
with the result obtained in the isospin limit.  The isospin-limit value
corresponds to each charge channel; the total $D\bar D$ width in this
limit would be twice this value.
}
\label{tab:isospin4040}
\begin{ruledtabular}
\begin{tabular}{cc}
Channel & Width (MeV) \\
\hline
$D^0\bar D^0$ & $(0.0101\pm0.0009)$ \\
$D^+D^-$ & $(0.0182\pm0.0016)$ \\
isospin limit, each charge channel & $(2.71\pm0.24)\times10^{-4}$ \\
\end{tabular}
\end{ruledtabular}
\end{table}

As expected, once the isospin limit is imposed, the two charge channels
become degenerate. More importantly, the common isospin-limit width is
even smaller than both physical charge-channel widths. This may appear
counterintuitive, since one would normally expect isospin breaking to be
a small correction. The explanation is that the $D\bar D$ amplitude is
already near zero. In this regime, small shifts of the kinematics or
wave function can move the overlap amplitude either closer to or farther
from the nodal zero. In the present case, the isospin-breaking inputs
happen to move the amplitude away from the zero and therefore increase
the width, whereas the isospin-limit kinematics place it even closer to
the cancellation point. The charged-neutral difference should therefore
be understood as an amplification of small isospin-breaking inputs near
a nearly vanishing amplitude, rather than as a signal of strong isospin
violation. This provides a further consistency check of the nodal
mechanism proposed here.

\subsection{Robustness of the mechanism}

The above results establish a coherent picture of the open-charm decays of $\psi(4040)$ within the present framework. The overall hierarchy follows naturally from the relativistic BS wave functions and the $^3P_0$ transition operator with a decay strength fixed independently from $\psi(3770)\to D\bar D$. In this picture, the strongly suppressed $D\bar D$ mode is traced to a channel-dependent cancellation induced by the nodal structure of the $3S$ wave function, while the sharp mass dependence and the charged-neutral sensitivity appear as secondary signatures of the same mechanism.

At the same time, the comparison between $M_A=4040$ MeV and $M_A=4051$ MeV makes clear that the absolute value of the $D\bar D$ width is numerically fragile in the present potential model. Because the physical region lies close to a zero of the overlap amplitude, a mass shift of only $11$ MeV can change the nearly vanishing $D\bar D$ width substantially, even though the qualitative hierarchy among the nonsuppressed channels remains stable. The present results therefore support the robustness of the suppression mechanism itself, rather than a precise prediction for the absolute $D\bar D$ partial width.

Taken together, these features indicate that the suppression of $\psi(4040)\to D\bar D$ is not a numerical accident within the present BS+$^3P_0$ treatment. Rather, it can already be understood as a dynamical consequence of the relativistic $3\,{}^3S_1$ Bethe--Salpeter wave function. Additional effects beyond the present framework, such as state mixing or coupled-channel dynamics, may still modify the quantitative widths, but they are not required in order to account for the basic channel hierarchy emphasized here.

\section{Discussion and conclusions}

In this work, we have studied the suppressed $D\bar D$ mode of $\psi(4040)$ and the associated open-charm hierarchy in the instantaneous Bethe--Salpeter framework combined with the relativistic $^3P_0$ model. The main result is that the characteristic hierarchy
\begin{equation}
\Gamma(D\bar D^*+{\rm c.c.}) \gg \Gamma(D_s\bar D_s) \gg \Gamma(D\bar D)
\end{equation}
arises naturally for a predominantly $3\,{}^3S_1$ charmonium assignment, with the suppressed $D\bar D$ mode traced to a sign-changing cancellation in the full relativistic overlap amplitude.

This point sharpens the usual qualitative nodal interpretation. In the BS framework, the node is distributed among several Salpeter components, so the physically relevant object is not any single radial function but the complete channel-dependent overlap integrand. The analysis of the momentum-dependent integrands and cumulative integrals shows explicitly that, for $\psi(4040)\to D\bar D$, positive and negative momentum regions contribute with comparable size and largely cancel. By contrast, $D\bar D^*$ and $D_s\bar D_s$ probe neighboring momentum regions in which the same initial-state wave function does not produce an equally strong cancellation. For the pseudoscalar-vector channels, the two structures ${\cal M}_{12}$ and ${\cal M}_{21}$ enter the width through $|{\cal M}_{12}|^2+|{\cal M}_{21}|^2$, so opposite signs of the amplitudes do not lead to a comparable suppression of the physical rate.

The mass-scan analysis provides an additional consistency check of this interpretation. Using a fixed nearby $3S$ wave function, we find that the physical region lies close to a zero of the $D\bar D$ overlap amplitude: the $D\bar D$ width develops a pronounced dip, whereas the $D_s\bar D_s$ and $D\bar D^*+{\rm c.c.}$ widths remain smooth and of similar order of magnitude. Likewise, the charged-neutral difference in the $D\bar D$ widths is naturally amplified because the amplitude already lies near a nodal zero. These correlated features support the interpretation that the observed suppression is channel selective and originates from the structure of the relativistic overlap amplitude.

An important aspect of the present analysis is that the pair-creation strength is not tuned to the $\psi(4040)$ data, but fixed independently from $\psi(3770)\to D\bar D$. In this sense, the suppressed $D\bar D$ mode and the associated open-charm hierarchy emerge as nontrivial consequences of the relativistic wave functions and decay overlaps, rather than as the result of a direct fit to the $\psi(4040)$ decays considered here.

The present study also has clear limitations. First, the calculation is restricted to the three channels $D\bar D$, $D\bar D^*+{\rm c.c.}$, and $D_s\bar D_s$, which isolate the core hierarchy relevant to the $\psi(4040)$ puzzle, but do not provide a complete description of the total open-charm width. In particular, the experimentally important $D^*\bar D^*$ channel is not included. Second, because the physical region lies close to a zero of the $D\bar D$ amplitude, the absolute value of the $D\bar D$ partial width is sensitive to the $11$ MeV mismatch between the BS eigenmass and the experimental pole. The quantitative prediction for this near-vanishing channel should therefore be viewed with appropriate caution, even though the qualitative suppression pattern, dip structure, and channel selectivity remain stable within the present analysis.

Within these limitations, the present results provide a transparent dynamical explanation of the suppressed $D\bar D$ mode and the core open-charm hierarchy of $\psi(4040)$ in a conventional quarkonium picture. More broadly, they illustrate that the open-flavor decays of radially excited quarkonia are controlled not only by phase space and spin counting, but also by the sign structure of relativistic overlap integrals. In this sense, $\psi(4040)$ offers a particularly clear example of how nodal filtering can shape the open-charm decays of excited charmonium.

\vspace{.5cm} {\bf Acknowledgments} \vspace{.5cm}

This work was supported by the National Natural Science Foundation of China under Grants No.~12575106, 12147214, and 12547114, and by the Specific Fund of Fundamental Scientific Research Operating Expenses for Undergraduate Universities in Liaoning Province under Grant No.~LJ212410165019.
The authors acknowledge the use of ChatGPT (OpenAI) and Aether (https://aether.aiphys.cn/) for language editing and presentation assistance. The scientific content, calculations, and conclusions are entirely the responsibility of the authors.

\end{document}